# Additivity of Quadrupole Moments in Superdeformed Bands: Single-Particle Motion at Extreme Conditions

W. Satuła[1−3], J. Dobaczewski[3], J. Dudek[4], W. Nazarewicz[2,3,5]

[1]*Joint Institute for Heavy Ion Research, Oak Ridge National Laboratory, P.O. Box 2008, Oak Ridge, TN 37831, U.S.A.*

[2]*Department of Physics, University of Tennessee, Knoxville, TN 37996, U.S.A.*

[3]*Institute of Theoretical Physics, Warsaw University, ul. Hoża 69, PL-00681, Warsaw, Poland*

[4]*Centre de Recherches Nucléaires, $IN_2P_3 - CNRS$/Université Louis Pasteur, F-67037 Strasbourg Cedex 2, France*

[5]*Physics Division, Oak Ridge National Laboratory, P.O. Box 2008, Oak Ridge, TN 37831, U.S.A.*

## Abstract

Quadrupole and hexadecapole moments of superdeformed bands in the $A\sim150$ mass region have been analyzed in the cranking Skyrme-Hartree-Fock model. It is demonstrated that, independently of the intrinsic configuration and of the proton and neutron numbers, the charge moments calculated with respect to the doubly-magic superdeformed core of $^{152}$Dy can be expressed very precisely in terms of independent contributions from the individual hole and particle orbitals. This result, together with earlier studies of the moments of inertia distributions, suggests that many features of the superdeformed bands in the $A\sim150$ mass region can be very well understood in terms of an almost undisturbed single-particle motion.

Typeset using REVTEX

Superdeformed (SD) nuclei in the $A\sim150$ mass region can be viewed as unique laboratories of nuclear structure. In these nuclei, due to strong shell effects and rapid rotation, residual pairing correlations are predicted to be weak. Consequently, the SD nuclei in the vicinity of $^{152}$Dy offer the possibility to study various facets of the single-particle motion in a rotating super-prolate well. According to the mean-field theory, the origin of SD bands can be traced back to large shell gaps in the single-particle spectra appearing simultanously for both protons and neutrons at large deformations. In the $A\sim150$ mass region, theory predicts that the region of strong SD shell effects spans over a relatively broad range of quadrupole deformations with the ($Z$=66, $N$=86)-system, $^{152}$Dy, being a doubly-magic SD nucleus [1–4].

The intrinsic configurations of SD states around $^{152}$Dy are well characterized by the intruder orbitals carrying large principal oscillator numbers $\mathcal{N}$ (high-$\mathcal{N}$ orbitals) [5–7]. These are the proton $\mathcal{N}$=6 and neutron $\mathcal{N}$=7 states. Because of their large intrinsic angular momenta and quadrupole moments, high-$\mathcal{N}$ orbitals strongly respond to the Coriolis interaction and to the deformed average field. Consequently, their occupation numbers are very good characteristics of rotational and deformation properties of SD bands, such as the dynamic moment of inertia, $\mathcal{J}^{(2)}$, or the transition quadrupole moment, $Q_t$. In particular, studies of the $\mathcal{J}^{(2)}$-behavior as a function of rotational frequency and the analysis of distributions of fractional changes in the moments of inertia [8] suggest that SD bands in the $A\sim150$ mass region can be well classified in terms of the high-$\mathcal{N}$ content, $\pi 6^n \nu 7^m$. The structural differences between SD bands possessing the same intruder content are usually small and they are due to other (non-intruder) states.

Recent progress in experimental techniques makes it possible to perform very precise relative lifetime measurements of yrast and excited SD bands using the Doppler shift attenuation method [9–11]. The transition quadrupole moments determined from experimental lifetimes are very valuable sources of information on shapes of SD bands and on intrinsic quadrupole moments and polarizabilities carried out by single particles and holes around the doubly-magic SD $^{152}$Dy core. According to theory, the intrinsic quadrupole moments are very robust observables; they are much less sensitive to the details of single-particle spectra (such as small shifts of single-particle energies, band crossings, etc.) and to rotation as compared to the moments of inertia. Therefore, they are very strong fingerprints of intrinsic single-particle configurations allowing for the stringent verification of theoretical configuration assignments.

In the present work, we performed a systematic theoretical analysis of quadrupole and hexadecapole moments in SD bands around $^{152}$Dy using the cranking Hartree-Fock (HF) Skyrme model (without pairing) of Ref. [12]. In the self-consistent approach, time-odd components of the rotating mean field appear naturally; these fields have been found to be important for the description of identical bands and moments of inertia [12,13]. All details of calculations strictly follow Ref. [12]. In order to demonstrate the sensitivity of results to the effective interaction, two different Skyrme parametrizations have been used: SkM* [14] and SkP [15].

The calculations have been carried out for the $N$=84-87 isotones of Gd, Tb, Dy, and Ho, for a number of (many)particle -(many)hole excitations with respect to the lowest SD band in $^{152}$Dy. The intrinsic single-particle orbitals have been labeled by means of the corresponding asymptotic quantum numbers characterizing the dominant oscillator component in the HF wave function. Specifically, the $\nu[770]$, $\nu[651]$, $\nu[642]$, $\nu[411]$, $\pi[651]$, $\pi[301]$ hole orbitals, and $\nu[761]$, $\nu[402]$, $\nu[514]$, $\nu[521]$, $\pi[530]$, $\pi[404]$, $\pi[411]$ particle orbitals have been considered. The corresponding Nilsson diagram can be found, e.g., in Ref. [6].

The calculated charge quadrupole moments, $Q_2$, in the SD nuclei in question decrease gradually with rotational frequency [4–6]. However, in a frequency range from $\hbar\omega$=0.2 MeV ($I\approx 20\,\hbar$) to $\hbar\omega$=0.5 MeV ($I\approx 50\,\hbar$) the relative variation of $Q_2$ is much smaller than 1% for most of the calculated bands. Therefore, in order to simplify the presentation, the $\omega$-averaged moments have been extracted from the calculations. (It is worth noting that the experimental values of $Q_t$ are also averaged over many transition energies.)

According to our HF calculations, namely those of Ref. [16], and the cranked relativistic mean-field approach [17], theoretical intrinsic charge quadrupole moments $Q_2$ are systematically larger by about 5% as compared to the experimental transition moments $Q_t$. For example, the experimental and theoretical values of the quadrupole moments in the yrast SD band in $^{152}$Dy are $Q_t$=17.5(2) eb [10] and $Q_2$=18.49 eb (HF+SkP), respectively. For the $\pi 6^2 \nu 7^1$ band in $^{149}$Gd, the experimental value is $Q_t$=15.0(2) eb [10] while theoretically $Q_2$=16.07 eb (HF+SkP). Because the deformations are large, one may, in principle, safely identify the intrinsic and transition moments to better than 5%. The origin of the obtained discrepancy is yet to be understood, although, as discussed in Ref. [9], experimental absolute values of $Q_t$ are, e.g., subject to a (10-15)% systematic error from uncertainties in the stopping powers.

Interestingly, the measured *relative* variations in $Q_t$ between SD bands are well reproduced by calculations. For instance, for the pair of SD bands discussed above, the difference $Q_t(^{149}\text{Gd};\pi 6^2\nu 7^1)$–$Q_t(^{152}\text{Dy;yrast})$ is $-2.5(3)$ eb in experiment while the predicted values are $-2.42$ eb and $-2.32$ eb in HF+SkP and HF+SkM*, respectively. Consequently, in the following discussion we concentrate on the calculated *relative intrinsic charge multipole moments* with respect to those of the yrast SD band in $^{152}$Dy:

$$\delta Q_\lambda(^A Z; c) \equiv Q_\lambda(^A Z; c) - Q_\lambda(^{152}\text{Dy; yrast}), \tag{1}$$

where $c$ stands for the configuration of the SD band in the nucleus $^A Z$, and $\lambda$=2 or 4 for the quadrupole or hexadecapole moments, respectively.

Figure 1 displays the relative charge quadrupole moments $\delta Q_2$ versus the relative charge hexadecapole moments $\delta Q_4$ for 74 SD bands calculated in the HF+SkP model. The $\delta Q_2$ values range from almost $-3$ eb to almost 1 eb, and most of the calculated points tend to correlate along a straight line.

The detailed analysis of results presented in Fig. 1 suggests that the multipole moments $\delta Q_2$ and $\delta Q_4$ obtained in the self-consistent HF calculations can be expressed as sums of individual contributions $q_\lambda(i)$ induced by individual particle and hole states. Namely, the relative charge multipole moments can be very well approximated by the "extreme shell model" expression

$$\delta Q_\lambda \approx \delta Q_\lambda^{\text{SM}} = \sum_i q_\lambda(i), \tag{2}$$

where $i$ runs over the particles and holes with respect to the ($^{152}$Dy; yrast) core defining the intrinsic SD configuration in the nucleus $^A Z$.

The quantity $q_\lambda(i)$ represents the effective single-particle multipole moment, i.e., the change of the total intrinsic moment which is induced on the whole nucleus by the given particle or hole. Following the standard shell model language, $q_\lambda(i)$ can be written as

$$q_\lambda(i) = q_\lambda^{\text{bare}}(i) + q_\lambda^{\text{pol}}(i), \qquad (3)$$

where $q_\lambda^{\text{bare}}(i)$ is the bare single-particle moment (in our calculations it is the single-particle HF charge multipole moment in $^{152}$Dy) and $q_\lambda^{\text{pol}}(i)$ represents the contribution due to the core polarization. Since the neutrons do not carry electric charge, their $q_\lambda^{\text{bare}}(i){=}0$, and the effective neutron moments come entirely from the core polarization effect entering through the self-consistency of the nuclear mean field. The additivity of single-particle multipole moments [Eq. (2)] means that the polarizations due to different particles and/or holes are to a large extent independent of one another.

Since the single-particle content of the HF bands is known, the effective single-particle multipole moments can be easily extracted from the calculated values of $\delta Q_\lambda$. This has been done in two steps. Firstly, by considering the SD bands differing from the doubly-magic core of $^{152}$Dy only by high-$\mathcal{N}$ particles or holes, the values of $q_\lambda$ have been extracted for the intruder orbitals. Next, by subtracting the contributions from high-$\mathcal{N}$ states, the values of $q_\lambda$ were found for non-intruder states by means of the least-squares fit. The calculated effective single-particle quadrupole moments are collected in Table I for both HF+SkP and HF+SkM$^*$ models. These numbers allow for a very precise determination of the total charge quadrupole or hexadecapole moment of a SD band calculated in HF, viewed as a (multi)particle-(multi)hole state with respect to the doubly-magic SD core of $^{152}$Dy.

The quality of the extreme single-particle picture is shown in the insert in Fig. 1, where we plot for the same set of SD bands the differences between the self-consistent values $\delta Q_\lambda$ and the shell-model values $\delta Q_\lambda^{\text{SM}}$ given by Eq. (2). The vast majority of calculated points falls into a very narrow region with $|\delta Q_2 - \delta Q_2^{\text{SM}}| \leq 0.05\,\text{eb}$ and $|\delta Q_4 - \delta Q_4^{\text{SM}}| \leq 0.03\,\text{eb}^2$. In a few cases (shown by the full dots) a very strong mixing between single-particle orbitals carrying very different single-particle quadrupole and hexadecapole moments gives rise to slightly larger deviations, but even in those situations the largest departure from the shell-model relation is around $\pm 0.2\,\text{eb}$ for $\delta Q_2$.

According to Table I, both Skyrme parametrizations employed give rather similar values of $q_\lambda$. The absolute values of effective single-particle multipole moments are systematically smaller in SkM$^*$ than in SkP, although these differences are of the order of the theoretical uncertainties due, e.g., to basis optimization and the statistical error of the least-squares procedure. For $q_2$ these uncertainties can be estimated to be of the order of 0.05 eb.

The calculated and experimental values of $\delta Q_2$ are compared in Table II. The experimental error bars in $\delta Q_2$ have been estimated under the assumption that the quoted experimental errors in $Q_t$ are statistical. The configuration assignments follow those of Refs. [10,11] except for bands 2 and 3 in $^{151}$Dy. It is seen that the general agreement between experiment and theory is good. The only serious discrepancy is seen for band 3 in $^{149}$Gd. As far as bands 2 and 3 in $^{151}$Dy are concerned, the experiment [11] yields positive values of $\delta Q_2$, and this makes it difficult to explain their structure in terms of a single-neutron hole in $\nu[642]^{-1}$ or $\nu[651]^{-1}$ states, both carrying negative quadrupole moments (Table I). On the other hand, the $\nu[411]^{-1}$ hole is consistent with the positive value of $\delta Q_2$. In this context, it is interesting to note that the particle-hole excitation $\pi[301]_+^{-1}\pi[530]_+$ has a large quadrupole moment of

~0.7 eb (see Table I). (The $\pi[530]_+$ routhian appears just above the $Z=66$ gap in the SkM* calculations.) Guided by the experimental values of $\delta Q_2$, we propose assignments shown in Table II.

The state dependence of effective single-particle quadrupole moments can be understood using the axial harmonic oscillator (HO) model. By applying the self-consistency condition for the HO [1], one arrives at the expression for the relative charge quadrupole moment [18]:

$$\delta Q_2^{\text{HO}} \approx A_{\text{HO}}^{\text{pol}}(\delta n_z^\nu + \delta n_z^\pi) + B_{\text{HO}}^{\text{pol}}(\delta n_\perp^\nu + \delta n_\perp^\pi) \\ + A_{\text{HO}}^{\text{bare}}\delta n_z^\pi + B_{\text{HO}}^{\text{bare}}\delta n_\perp^\pi, \tag{4}$$

where $\delta n_z$ and $\delta n_\perp$ represent the changes in the numbers of oscillator quanta brought into the system by the particle and/or hole excitations with respect to the $^{152}$Dy core. In analogy to Eq. (3), the coefficients $A^{\text{pol}}$ and $B^{\text{pol}}$ represent the change in $Q_2$ due to the core polarization, while $A^{\text{bare}}$ and $B^{\text{bare}}$ represent the direct proton contributions. For the SD core $Z=60$ and $N=80$, the coefficients in Eq. (4) are: $A_{\text{HO}}^{\text{pol}}=0.051$ eb, $B_{\text{HO}}^{\text{pol}}=-0.045$ eb, $A_{\text{HO}}^{\text{bare}}=0.173$ eb, and $B_{\text{HO}}^{\text{bare}}=-0.030$ eb.

The least-squares fit to the HF+SkP values of $\delta Q_2$ yields the corresponding HF values: $A_{\text{HF}}^{\text{pol}}=0.092$ eb, $B_{HF}^{\text{pol}}=-0.087$ eb, $A_{\text{HF}}^{\text{bare}}=0.116$ eb, and $B_{\text{HF}}^{\text{bare}}=0.031$ eb. By comparing the HO and HF results, one can see that the quadrupole polarizability in HO is roughly two times smaller than in the HF model. The reason for this discrepancy is well known: the $\Delta \mathcal{N}=2$ coupling to the giant quadrupole resonance is absent in the simple HO description. In Ref. [18] we discuss the core polarization effects within the RPA+HO framework using the doubly-stretched quadrupole-quadrupole interaction [19]. (For a comparison between HO and HF quadrupole polarizabilities at ground-state deformations, see Ref. [20].)

The HO expression [Eq. (4)] nicely illuminates the self-consistent results for $q_2$ shown in Table I. In particular, it explains the dominant role of high-$\mathcal{N}$ orbitals in terms of their large values of $n_z$.

The strong configuration dependence of intrinsic quadrupole moments have been previosly discussed in the cranked Nilsson-Strutinsky calculations of Ref. [10] where the quadrupole moments were extracted from the calculated equilibrium deformations assuming a sharp uniform charge distribution. Since the macroscopic quadrupole moments scale according to $ZA^{2/3}$, the authors concluded that the equality of quadrupole moments in different nuclei must necessarily imply different equlibrium deformations. This conclusion is not supported by our calculations. Similar quadrupole moments in identical SD bands can be simply explained by their identical high-$\mathcal{N}$ content. (As seen in Table I the high-$\mathcal{N}$ intruder states carry the largest values of $q_\lambda$.) The remaining differences can be attributed to other valence orbitals. The macroscopic $ZA^{2/3}$ scaling does not hold in microscopic theory; it is the identity of quadrupole moments (which, after all, are observables), not the difference in quadrupole deformations, that is vital for understanding the identical band phenomenon.

In summary, we performed a systematic study of quadrupole and hexadecapole moments in SD bands in the $A \sim 150$ mass region using the Skyrme-HF model. The most striking observation is that the relative quadrupole moments can be written as a sum of independent contributions from the single-particle/hole states around the doubly-magic SD core of $^{152}$Dy. That is, the "extreme shell-model" relation [Eq. (2)] holds with a surprisingly high accuracy. The relative quadrupole moments calculated with respect to the $^{152}$Dy SD core follow experimental trends rather well, and they are rather strong indicators of underlying intrinsic configurations. Our results, together with the previous systematics of experimental moments of inertia [8], strongly suggest that the SD high-spin bands around $^{152}$Dy are excellent examples of an almost undisturbed single-particle motion.

## ACKNOWLEDGMENTS

Useful discussions with R.V.F. Janssens, E.F. Moore, D. Nisius, and B. Haas are gratefully acknowledged. This research was supported in part by the U.S. Department of Energy (DOE) through Contract Nos. DE-FG05-93ER40770 (University of Tennessee) and DE-FG05-87ER40361 (Joint Institute for Heavy Ion Research), and by the Polish Committee for Scientific Research under Contract No. 2 P03B 034 08. Oak Ridge National Laboratory is managed for the U.S. DOE by Lockheed Martin Energy Research Corp. under Contract No. DE-AC05-96OR22464. We would like to express our thanks to the *Institut du Développement et de Ressources en Informatique Scientifique* (IDRIS) of CNRS, France, which provided us with the computing facilities under Project No. 940333.

TABLES

TABLE I. Effective charge quadrupole, $q_2$, and hexadecapole, $q_4$, moments for single-hole and single-particle orbitals around the $^{152}$Dy core. The values were extracted from the set of 74 calculated bands in HF+SkP and 133 bands in HF+SkM*. The orbitals are labeled by means of the asymptotic quantum numbers $[Nn_z\Lambda]$ corresponding to the dominant harmonic oscillator component in the HF wave function and the signature quantum number $r$ (the subscripts $\pm$ stand for $r=\pm i$). Note the very weak signature-dependence of $q_\lambda$.

| $[Nn_z\Lambda]_r$ | $q_2$ (eb) | | $q_4$ (eb$^2$) | |
| --- | --- | --- | --- | --- |
| | SkP | SkM* | SkP | SkM* |
| | | holes | | |
| $\pi[651]_+$ | −0.96 | −0.96 | −0.22 | −0.22 |
| $\pi[651]_-$ | −0.89 | −0.88 | −0.19 | −0.19 |
| $\pi[301]_+$ | 0.15 | 0.13 | −0.01 | −0.01 |
| $\pi[301]_-$ | 0.18 | 0.16 | −0.01 | −0.01 |
| $\nu[770]_+$ | −0.59 | −0.48 | −0.17 | −0.17 |
| $\nu[770]_-$ | −0.57 | −0.48 | −0.17 | −0.17 |
| $\nu[642]_+$ | −0.22 | −0.22 | −0.03 | −0.03 |
| $\nu[642]_-$ | −0.24 | −0.24 | −0.03 | −0.03 |
| $\nu[651]_+$ | −0.43 | −0.28 | −0.09 | −0.05 |
| $\nu[651]_-$ | −0.43 | −0.30 | −0.09 | −0.05 |
| $\nu[411]_+$ [a] | 0.18 | 0.16 | 0.05 | 0.05 |
| $\nu[411]_-$ [a] | 0.15 | 0.13 | 0.05 | 0.05 |
| | | particles | | |
| $\pi[530]_{+/-}$ | 0.59 | 0.55 | 0.04 | 0.03 |
| $\pi[411]_{+/-}$ | 0.11 | 0.10 | −0.05 | −0.04 |
| $\pi[404]_{+/-}$ | −0.30 | −0.28 | −0.01 | 0.00 |
| $\nu[402]_{+/-}$ | −0.44 | −0.38 | −0.03 | −0.02 |
| $\nu[514]_{+/-}$ | −0.25 | −0.22 | −0.07 | −0.06 |
| $\nu[521]_{+/-}$ | 0.00 | −0.01 | −0.03 | −0.03 |
| $\nu[761]_+$ | 0.41 | 0.28 | 0.09 | 0.06 |
| $\nu[761]_-$ | 0.46 | 0.41 | 0.11 | 0.11 |

[a] In the HF+SkP model this orbital should be labeled as [413] rather than [411].

TABLE II. Experimental and calculated (HF+SkP and HF+SkM*) relative charge quadrupole moments $\delta Q_2$, Eq. (1), in SD bands in $^{148,149}$Gd, $^{151}$Tb, and $^{151}$Dy.

| Nucleus | configuration | $\delta Q_2$ (eb) | | |
|---|---|---|---|---|
| | | EXP | SkP | SkM* |
| $^{148}$Gd | $\nu[770]_-^{-1}\nu[651]_+^{-1}\pi[651]^{-2}$ | $-2.9(0.3)^a$ | $-2.85$ | $-2.60$ |
| | $\nu[651]^{-2}\pi 6^{-2}$ | $-2.7(0.4)^a$ | $-2.71$ | $-2.42$ |
| | $\nu[411]^{-2}\pi[301]^{-2}$ | $0.3(1.3)^a$ | $0.66$ | $0.58$ |
| $^{149}$Gd | $\nu[770]_-^{-1}\pi[651]^{-2}$ | $-2.5(0.3)^a$ | $-2.42$ | $-2.32$ |
| | $\nu[651]_-^{-1}\pi[651]^{-2}$ | $-1.9(0.4)^a$ | $-2.28$ | $-2.12$ |
| | $\nu[770]_-^{-1}\pi[651]_+^{-1}\pi[301]_+^{-1}$ | $-2.3(0.5)^a$ | $-1.38$ | $-1.31$ |
| | $\nu[411]_-^{-1}\pi[301]^{-2}$ | $0.0(0.6)^a$ | $0.48$ | $0.42$ |
| $^{151}$Tb | $\pi[651]_+^{-1}$ | $-0.7(0.7)^b$ | $-0.96$ | $-0.96$ |
| $^{151}$Dy | $\nu[770]_-^{-1}$ | $-0.6(0.4)^b$ | $-0.57$ | $-0.48$ |
| | $\nu[642]_+^{-1}\pi[301]_+^{-1}\pi[530]_+^{c}$ | $0.7(0.7)^b$ | $0.52$ | $0.46$ |
| | $\nu[411]_+^{-1c}$ | $0.4(0.6)^b$ | $0.18$ | $0.16$ |
| | $\nu[411]_-^{-1}$ | $0.0(0.6)^b$ | $0.15$ | $0.13$ |

$^a$Ref. [10]. $^b$Ref. [11]. $^c$See discussion in text.

FIGURES

FIG. 1. Relative charge quadrupole moments $\delta Q_2$, versus relative charge hexadecapole moments, $\delta Q_4$, Eq. (1), for 74 SD bands calculated in the HF+SkP model. The insert (in the same scale) shows the differences $\delta Q_\lambda - \delta Q_\lambda^{\rm SM}$, Eq. (2).

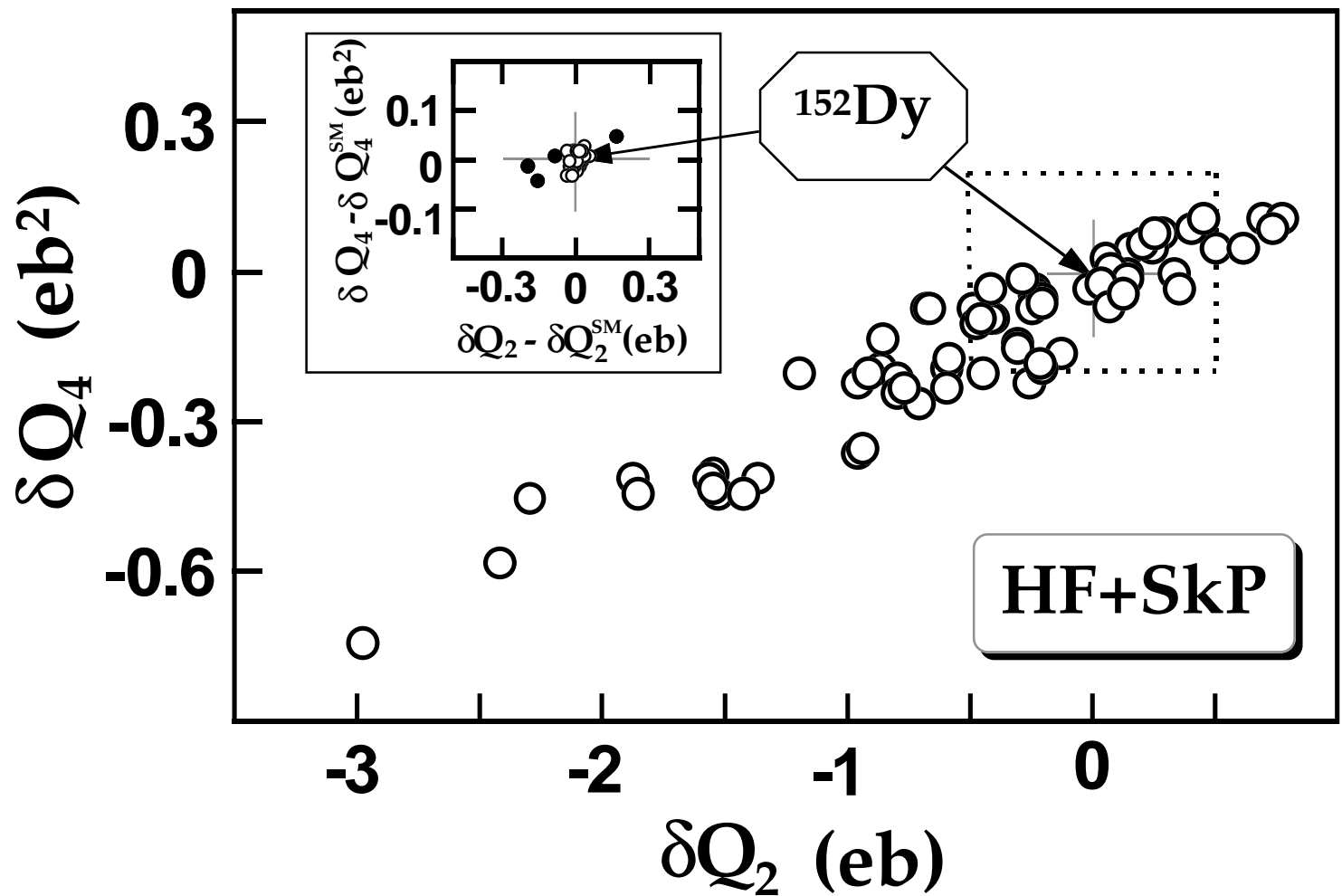

Figure. 1